# Research progress of rubrene as an excellent multifunctional organic semiconductor

Si Liu,† Hongnan Wu,† Xiaotao Zhang, * Wenping Hu

Tianjin Key Laboratory of Molecular Optoelectronic Science, Department of Chemistry, School of Science, Tianjin University & Collaborative Innovation Center of Chemical Science and Engineering (Tianjin), Tianjin 300072, China.

E-mail: zhangxt@tju.edu.cn.

† These authors contributed equally to this work

Rubrene, a superstar in organic semiconductors, has achieved unprecedented achievements in the application of electronic devices, and research based on its various photoelectric properties is still in progress. In this review, we introduced the preparation of rubrene crystal, summarized the applications in organic optoelectronic devices with the latest research achievements based on rubrene semiconductors. An outlook of future research directions and challenges of rubrene semiconductor for applications is also provided.

## 1  Introduction

Organic semiconducting materials with the advantages low-cost, designability of structure, solution processability and flexibility[1-3] are widely used in organic electronic devices, including organic field-effect transistors (OFETs) , organic photovoltaics (OPVs), and organic light-emitting diodes(OLEDs), playing an important role in the field of organic optoelectronics[4-9]. Rubrene(5,6,11,12-tetraphenylnaphthacene, $C_{42}H_{28}$, **Fig. 1**), a p-type organic semiconductor, is a superstar in electronic device applications, exhibiting a high charge carrier mobility of 20 $cm^2\ V^{-1}\ s^{-1}$ at room temperature and a fluorescence quantum efficiency of approximately 100% in solution[10,11].

The structure of rubrene molecular is shown in the **Fig. 1(a)**, which is composed of a tetracene backbone and four phenyl groups substituted at the 5, 6, 11, 12 positions；its powder is red or light brown, with strong absorption in the blue-violet to green spectrum. Rubrene crystals have three polymorphs：orthorhombic, monoclinic, and triclinic, as shown in **Fig. 1(b-d)**. Only orthorhombic presents a herringbone molecular arrangement, large π-π orbital overlap of the tetracene backbone of the adjacent molecules, which is required for high-performance charge transfer[12-14]. The triclinic and monoclinic do not have this effective molecular stacking, so OFETs research focuses on the orthorhombic. Rubrene has highly delocalized conjugated π bond , enhancing the stability of the molecule and giving a low-energy light-emitting band, which is beneficial for application in OLEDs[15-17]. In addition, the singlet exciton split quantum yield of rubrene is close to 100%，attracting the applications of rubrene in OPVs[18-20].What's more, the properties of low spin -orbit coupling and weak hyperfine interactions of organic semiconductors , expands the application in the field of spintronics for rubrene[21,22]. Therefore, rubrene is a multifunctional organic semiconductor that combines various advantages. Unique molecular structure and photophysical properties make it shine in the application of electronic devices.

In this paper,we reviewed the applications of rubrene semiconductor from three parts: the preparation of rubrene crystal for devices, the latest research achievements and the development trend in the application of electronic devices based on rubrene semiconductors.

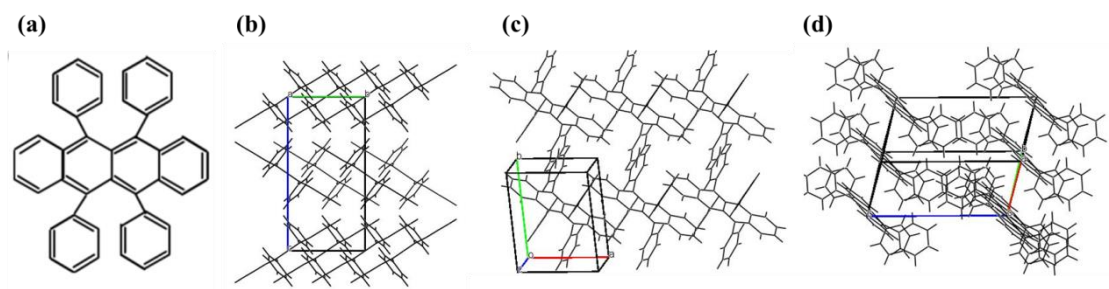

**Fig. 1** (a)Molecular structure of rubrene. Packing of rubrene in (b) orthorhombic, (c)triclinic and (d)monoclinic

## 2  Preparation

There are three crystal systems of rubrene, including orthorhombic, triclinic and monoclinic. The orthorhombic system of rubrene is usually prepared by physical vapor transport (PVT)[23-27]. Kim et al. used PVT method to obtain different size and thickness of rubrene single crystal by adjusting the crystallization time, carrier gas flow or source temperature[24]，(**Fig. 2(a)-(c)**). Based on this, they studied the relationship between the charge transport properties and the surface roughness of rubrene single crystal. With the increase of crystal thickness, the roughness increases gradually, and the surface conductivity decreases exponentially (crystal thickness is less than 4 μm). This paper shows that in order to understand the charge transport properties of a single crystal clearly, it is necessary to carefully consider the crystal morphology, such as thickness and roughness. And PVT method can be used to prepare different morphologies of rubrene crystals, such as needle-like, platelet, and so on[23,26]. Because PVT method is simple and convenient, it is widely used in rubrene single crystal devices. Orthorhombic crystal system can also be prepared by vacuum evaporation[28-29,13] and solution method[30-33]. Sim et al. used capillary tube-assisted solution process to directly grow rubrene crystals on substrate[33]. In this solution method, the solution evaporates from the edge of the droplet to the center. As the solvent volatilizes completely, rubrene will crystallize around the capillary tube, which can effectively control the growth position of the crystal on the polymer gate dielectric. At the same time, they also observed the growth of rubrene crystals in various concentrations (0.1, 0.01 and 0.001wt%). It is found that when the concentration is 0.01 wt%, the quality and properties of the crystal are the best. Based on this, the highest hole mobility of the OFET is 3.74 $cm^2\ V^{-1}\ s^{-1}$, and it has high stability under environmental conditions. The tube-assisted solution process developed by them is low cost, easy to operate and can well control the position of semiconductor crystals, which is helpful for large scale devices fabrication.

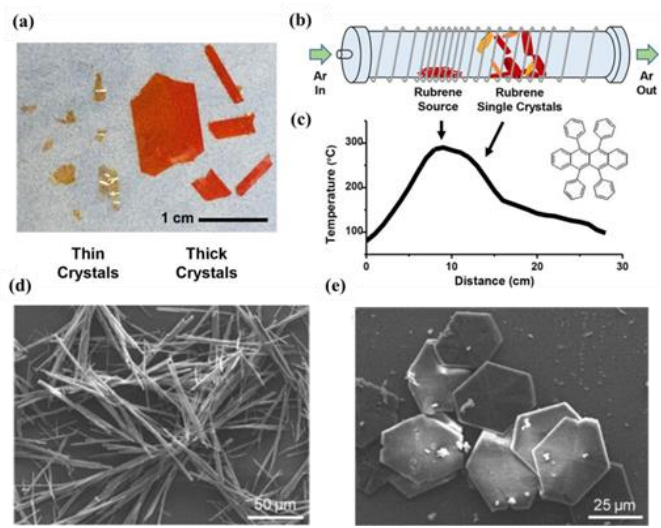

**Fig. 2** (a) Photos of different thicknesses of rubrene crystals. (b) Schematic diagram of crystal growth by PVT method. (c) The change of temperature in tube with distance. **Reproduced from Ref. [24]** SEM image of (d) 1D rubrene microwires and (e) 2D hexagonal plate. **Reproduced from Ref. [35]**

In the triclinic crystal system, the molecules are basically arranged in a parallel way in the plane, resulting in only a part of π orbital overlap in the stacking direction. This kind of crystal system can be made by solution method. For example, Matsukawa et al. grown two kinds of rubrene crystals, hexagons and parallelograms, by the solution-slow-cooling method with 1-propanol solvent, and they were orthorhombic and triclinic respectively[34]. At the same time, they measured the electrical properties of OFETs based on these two single crystals. Using the same OFET structure, the carrier mobility of triclinic crystal was significantly lower than that of orthorhombic crystal by an order of magnitude. Jo et al. obtained the blend films of different rubrene crystal structures (orthorhombic, triclinic) by solution-processing, annealing of rubrene and different polymer binder[31].

In monoclinic crystal system, two adjacent molecules are almost perpendicular to each other on the molecular plane, so there is no π-π accumulation between the molecules of rubrene in this crystal system. At present, there are few reports about the monoclinic system of rubrene. Huang et al. employed the reprecipitation method to obtain two kinds of rubrene crystals with different structures[12]. The specific operation is to inject 50 μL of 30 mM rubrene stock solution in chloroform into methanol, and by altering the volume of methanol, ribbons (triclinic), rhombic and hexagonal platesl (monoclinic) crystal are obtained. Their work has realized the controllable growth of different crystal forms of rubrene crystals, which has attracted many people's attention and reference. In 2017, Gu et al. used the same method to prepare 1D microwires (triclinic) and 2D hexagonal plates (monoclinic) crystals(**Fig. 2 (d)-(e)**) by changing the supersaturation of rubrene molecules[35]. These two microcrystals were used as active materials to construct electrogenerated chemiluminescence (ECL) sensors, and the sensors had excellent ECL responses to biological molecule in a wide linear range. The quantitative limits of triclinic crystal and monoclinic crysyal were $3 \times 10^{-14}$ M and $1 \times 10^{-13}$ M, respectively. Their report provides a new idea for the material research of ECL sensors. In the solution method, the types of solvents, concentration of rubrene and the temperature of crystal growth may all play an important role in which polymorph is formed. In addition to the above common methods for preparing rubrene single crystals, many novel methods have been developed in recent years, including epitaxy[36,37], microspating in air sublimation[38] and so on.

The crystallinity of rubrene also has a great influence on its performance. By introducing a modified layer[29,39-42], changing the substrate temperature[28,41], introducing doping[31] and so on, the crystallinity of rubrene can be effectively adjusted, and then the device performance can be optimized. In 2005, Haemori et al. prepared the crystalline thin film of rubrene with pentacene thickness gradient (0-1 monolayer) film as buffer layer by using the combinatorial molecular beam epitaxy[39]. Its on/off ratio and charge mobility were $10^6$ and 0.07 cm$^2$/v s, respectively. This is the pioneering work of the transformation of the rubrene film from amorphous to crystalline. Sun et al. used p-sexiphenyl (*p*-6P) and copper phthalocyanine (CuPc) as heterogeneous inducing bilayers, obtained the rubrene film by weak epitaxial growth[41]. After the introduction of heterogeneous inducing bilayer, the rubrene film changes to crystalline rubrene films, and the performance of the transistor is also improved significantly. They also studied the effect of the inducing bilayers on the film at different substrate temperatures (Tsub). When the substrate temperature is 90 ℃ and the thickness of the rubrene film is 20nm, the quality of the film is the best. Lin et al. studied the effect of Tsub (103 – 221℃) on the properties of rubrene films in vacuum deposition[28]. Through SEM, they found that the average size of rubrene grains increased with the increase of Tsub in a certain temperature range, and the shape of single crystal gradually changed from dendrite fibers to randomly distributed 3D microcrystalline. When Tsub was higher than 189℃ (lower than 221℃), thick films composed of large grains with high crystallinity will be formed. If Tsub continues to rise (above 221℃), the roughness of the crystal will increase to a large extent, and the growth rate of the crystal will also show a downward trend. Therefore, the temperature should be strictly controlled in the preparation of rubrene solid films for photoelectric devices by vacuum deposition.

## 3 Applications in electronic devices

### 3.1 Organic field-effect transistors(OFETs)

Organic field-effect transistors (OFETs) with advantages of low-cost and easy-fabrication, is the key element for realizing flexible and printed organic electronics, such as for amplifiers, electronic papers, active matrix displays, and radio frequency identification (RFID) cards[43].The field effect transistor based on rubrene single crystal has caused huge repercussions due to high hole carrier mobility, which has led to extensive research on rubrene[44-47] and has made great achievements for high performance transmission devices[33,38]. Here, we mainly summarized that the hole carrier mobility is above 10 cm$^2$ V$^{-1}$s$^{-1}$. Sundar et al. fabricated bottom-contact field-effect transistors of rubrene single crystal with polydimethylsiloxane (PDMS) as the substrate in order to avoid damage to the crystal by the device manufacturing process and systematically studied the anisotropic charge transport of the single crystal [48]. The charge mobility of rubrene along the b-axis direction is up to 15.4 cm$^2$ V$^{-1}$s$^{-1}$, and along the a-axis direction reaches 4.4 cm$^2$ V$^{-1}$s$^{-1}$. Takeya et al. achieved maximum transistor mobility of 18 cm$^2$ V$^{-1}$s$^{-1}$ and contact-free intrinsic value of 40 cm$^2$ V$^{-1}$s$^{-1}$ by using inner molecular layers of the semiconductor crystal for the conduction channel[49]. They also fabricated double gate transistors base on 1nm-thick rubrene crystals and 9,10-diphenylanthracene single crystal and surface-passivated $SiO_2$ were used for the gate insulators. High carrier mobility of 20–25 $cm^2 V^{-1}s^{-1}$ was obtained in response to the single gate voltages and 43$cm^2 V^{-1}s^{-1}$ was achieved for the double-gate rubrene transistor[50]. Zhang et al. studied the relationship between carrier mobility and single crystal length based on rubrene nanoribbon transistors. Maximum mobility for the device can reach 24.5 $cm^2 V^{-1}s^{-1}$[51]. They found that as the crystal widens and thickens, the mobility of the device decreases exponentially, but increases linearly with the increase of conductive channels at 3–15μm. Adhikari et al. reporeted a highest value of 12 $cm^2 V^{-1}s^{-1}$ for rubrene single crystals devices incorporating high-k (>3) polymer dielectrics by using crosslinked poly(vinylidene fluoride‑bromotrifluoroethylene) P(VDF‑BTFE) as the dielectric which controlled the chain conformations of gate insulator layers and reduced interfacial trap densities [52].

Rubrene not only exhibits excellent hole transport properties，but also achieve electron/ ambipolar transport by selecting appropriate dielectric layers and electrodes. [53-58]. Takahashi et al. firstly reported the ambipolar behavior in rubrene OFETs by using hydroxyl-free gate dielectric PMMA to reduce the electron traps[53]，**Fig. 3**. And later they improved the electron carrier mobility to 0.81 $cm^2 V^{-1}s^{-1}$ through multiple purifications of rubrene single crystal[55]. Kanagasekaran et al. proposed a new electrode structure made by metal layer (Ca or Au) /polycrystalline organic semiconductors(pc-OSC)/tetratetracontane(TTC) displaying high ambipolar injection efficiency, and they obtained the highest field-effect mobilities of holes (22 $cm^2 V^{-1} s^{-1}$) and electrons (5.0 $cm^2 V^{-1} s^{-1}$) for rubrene, **Fig. 4**[58]. In addition , rubrene single crystal FET with this new electrode structure showed a bright light emission with high current density of 25 kA $cm^{-2}$ and the device could work even after exposure to air, **Fig. 4(d)**.

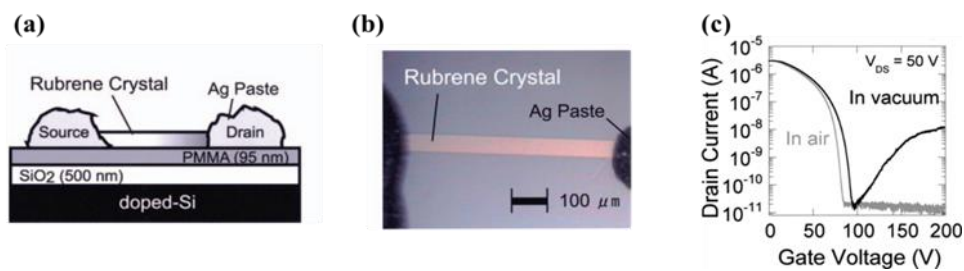

**Fig. 3** (a) Schematic illustration (b) optical image and (c) typical transfer curves of rubrene single-crystal OFET (black represents vacuum conditions and gray represents air atmosphere in (c). ) **Reproduced from Ref. [53]**

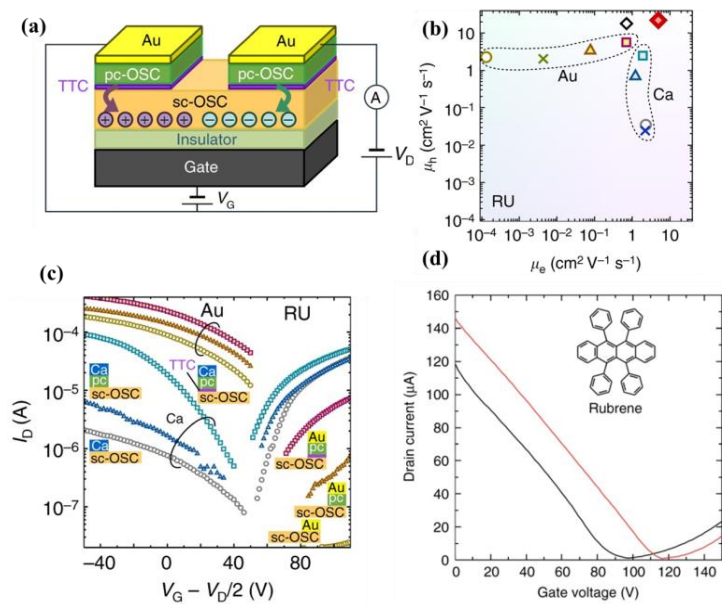

**Fig. 4** (a) A Schematic representation of a rubrene single crystal (sc-RU) FET with the new electrodes Au/pc-OSC/TTC. (b) Electron and hole mobilities and (c) transfer characteristics of sc-RU FET with different electrodes(Yellow-to-magenta and blue（gray）-to-cyanmarks indicate Au electrodes and Ca electrodes: Open circles, closed triangles, closed squares, and crosses indicate no interlayer, pc-OSC interlayer, pc-OSC/TTC interlayer, and CsF interlayer. Red and black diamonds indicate the highest mobilities of rubrene single-crystal FETs). (d) Transfer characteristics of the sc-Ru FET before (black) and after (red) air exposure of 2 h (The drain voltage $V_D$ was 150 V). **Reproduced from Ref. [58]**

In addition, as one of the most promising organic materials with high carrier mobility, rubrene single crystals are also often used to prepare heterojunctions to achieve balanced charge transport of electrons and holes in OFETs [59-62]. He et al. synthesized organic–inorganic $MoS_2$/rubrene FETs and achieved behavior with the well balanced electron and hole mobilities of 1.27 $cm^2\ V^{-1}\ s^{-1}$ and 0.36 $cm^2\ V^{-1}\ s^{-1}$ respectively, showing in **Fig. 5**[61]. The $MoS_2$/rubrene ambipolar transistors are used to fabricate CMOS (complementary metal oxide semiconductor) inverters that showed good performance with a gain of 2.3 at a switching threshold voltage of −26 V. Lately, A core-shell heterojunction fabricated by growing rubrene single crystals on N,N′-bis(2-phenylethyl)perylene-3,4:9,10-tetracarboxylic diimide (BPE-PTCDI) single crystal nanowires was reported by Lee et al. [62] (**see Fig. 6**). Distinct ambipolar charge transport behavior in the OFET, maximum electron and hole mobilities of $1.41 \times 10^{-1}$ and $8.31 \times 10^{-2}\ cm^2\ V^{-1}\ s^{-1}$ was observed and enabling the application of the core−shell crystal in complementary circuits. CMOS-like inverter based on these ambipolar OFETs showed sharp voltage conversion near 50 V at a supply voltage of 100 V with a gain of 23.7 in n-type operation mode.

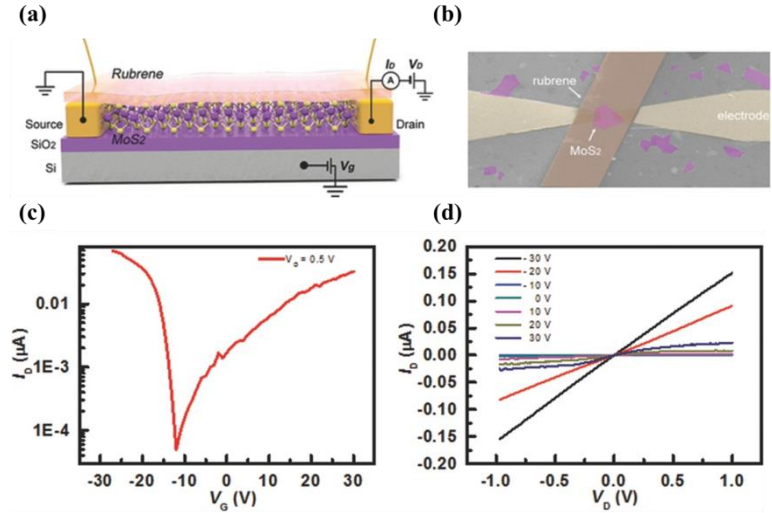

**Fig. 5** (a) Schematic illustration and (b) false-colored SEM image of the ambipolar FET based on the $MoS_2$-rubrene heterostructure. (c) The transfer and (d) output curves of the $MoS_2$-rubrene ambipolar FET. **Reproduced from Ref. [61]**

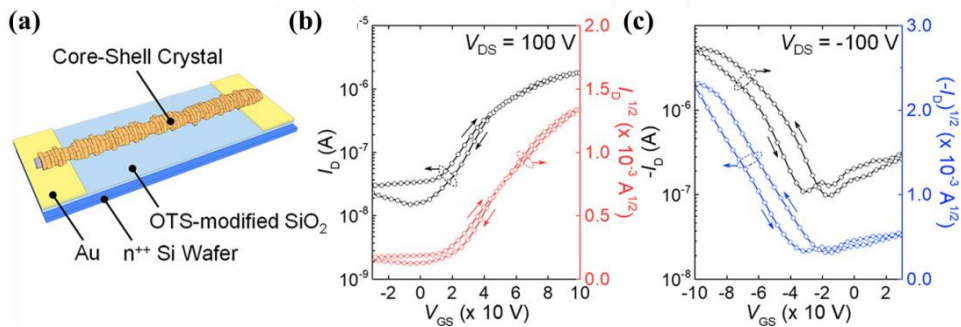

**Fig. 6** (a) Schematic illustration of the device structure based on ambipolar core-shell BPE-PTCDI/rubrene crystal. (b) The n-type and (c) p-type transfer characteristics of the BPE-PTCDI/rubrene crystal under dual sweeping mode. **Reproduced from Ref. [62]**

Single crystal devices of rubrene exhibit the highest charge transport performance, and the preparation of large-area crystalline thin films can expand further application in field effect transistors [23-26]. Hu et al. obtained highly ordered and uniform rubrene crystalline film using $C_{60}$(0.5nm) as a nucleation agent and p-6P(5nm) as a modification layer[67],**Fig. 7**. OFET devices fabricated by sequentially depositing p-6P,$C_{60}$(T=150 ℃, $C_{60}$ was deposited with substrate temperature of 150℃) and rubrene(10nm)on Si/$SiO_2$ substrate by vacuum evaporation method showed a high hole mobility of 1.0-1.4 $cm^2$ $V^{-1}$ $s^{-1}$ and a switching ratio of $10^6$. In addition, when $C_{60}$ was deposited on the substrate at 90℃, smoother rubrene film can be obtained with a higher mobility up to 2.95 $cm^2$ $V^{-1}$ $s^{-1}$. Fusella group discovered a thin (5 nm) organic underlayer between the substrate and rubrene could improve the crystallinity of rubrene thin films formed via postdeposition annealing[40]. Ultrathin, large-area, fully connected, and highly crystalline thin films of rubrene was obtained with highest hole carrier mobility of 3.5 $cm^2$ $V^{-1}$ $s^{-1}$, **Fig. 8**. The large-area crystallization technology breaks through the bottleneck of fabricating high mobility rubrene organic thin film transistor.

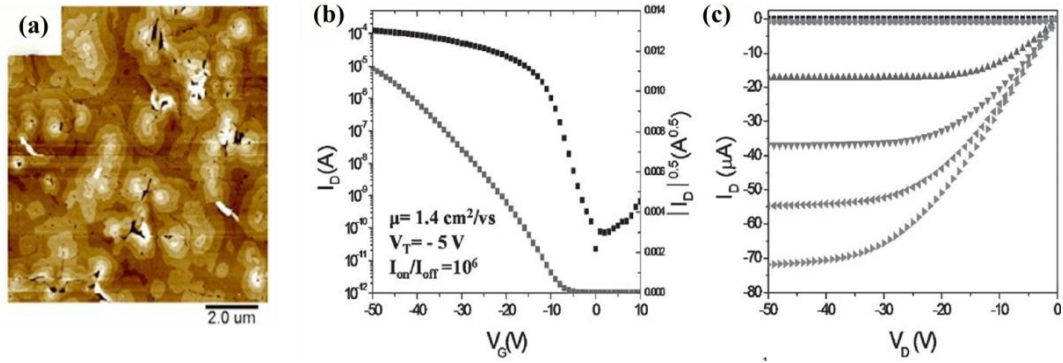

**Fig. 7** (a) AFM height image of 10 nm rubrene on 0.5 nm $C_{60}$/5 nm $p$-6P. (b) Transfer and (c) output characteristics of 10 nm rubrene/0.5 nm $C_{60}$ films (T=150℃). **Reproduced from Ref. [67]**

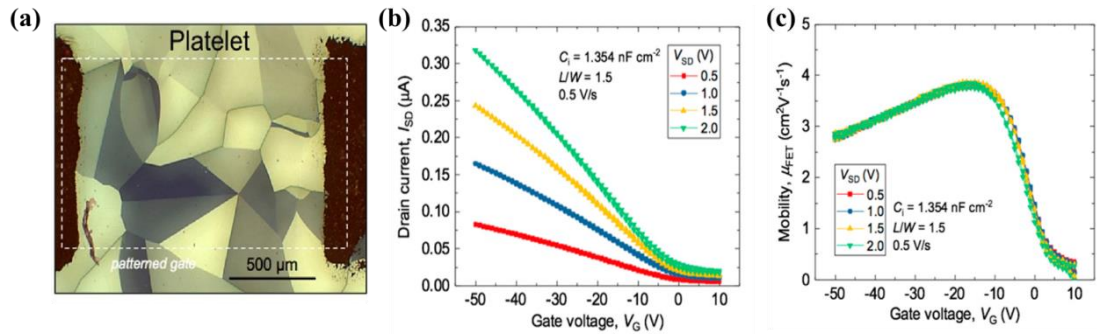

**Fig. 8** (a) Polarized optical microscope image of the rubrene films around the conducting channel. (b) Plots of the source/drain current ($I_{SD}$) and (c) FET mobilities in the linear regime ($\mu_{FET}$) as a function of gate voltage ($V_G$) for rubrene platelets. **Reproduced from Ref. [40]**

### 3.2 Organic light-emitting devices(OLEDs)

OLED have attracted more and more attention in display and energy-saving lighting technology due to their advantages of high EL efficiency, flexibility and low manufacturing cost. As an excellent organic semiconductor with efficient orange fluorescent, rubrene can be used as light-emitting layer[17,68,69]or dopant[70-74] to OLED. As a dopant, rubrene can effectively improve the luminous intensity and efficiency of OLED. Zhao et al. fabricated orange fluorescent OLED with structure of ITO/$MoO_3$ (3nm)/N,N'-bis-(1-naphthl)-diphenyl-1,1'-biphenyl-4,4'-diamine (NPB, 20 nm)/4,4',4''-tri(N-carbazolyl)triphenylamine (TCTA, 8 nm)/2,4,6-tris(3-(1H-pyrazol-1-yl)phenyl)-1,3,5-triazine (3P-T2T): x% rubrene (10 nm)/3P-T2T (30 nm)/LiF (1 nm)/Al which using doping rubrene[70]. And the influence of different concentration of rubrene (x=0, 1.0, 1.5, 2.0, 2.5 and 3.0%) doping on the performance of OLED was researched. When the concentration of rubrene is 1.5% the efficiency emission are the highest, the power efficiency, maximum current efficiency and external quantum effifiency (EQE) reach up to 22.6 lm/W, 25.3 cd/A and 8.1%, respectively. The reason for this phenomenon is attributed to high reverse intersystem crossing efficiency of triplet excitons and efficient energy transfer from thermally activated delayed flfluorescence exciplex to dopant. Li et al. proposed OLED with white light emission by doping rubrene[72]. The best chromaticity coordinate(0.319,0.317) of the device is very stable, and the highest efficiency and luminance show 5.1022 cd/A and 17130 cd/$m^2$, respectively.

Zhao et al. prepared a group of white OLED with rubrene as the yellow light layer and doping phosphorescent dye FIrPic(Bis(3,5-difluoro)-2-(2-pyridyl)phenyl-(2-carboxypyridyl)iridium(III)) in MCP(1,3-Bis(carbazol-9-yl)benzene) as the blue light layer, and optimized the device performance by changing the thickness of rubrene[68]. Compared with Alq3 as the electron transport layer (ETL), Zheng et al. showed a red fluorescent OLED using rubrene / bathophenanthroline (Bphen) as the ETL, and obtained higher EQE (4.67%) and the maximum current efficiency (5.50 cd/A，an increase of approximately 60%)[75]. Their work suggest that the performance of OLED can be improved by modifying hole injection layer and optimizing the emitting layer. Saikia et al. reported the first OLED with rubrene as buffer layer over the electrode surface, and the devices configuration are fluorine-doped tin oxide (FTO)/buffer layer(rubrene)/4,4'-bis[N-(3-methylphenyl)-N-phenylamino]biphenyl (TPD, 35 nm)/Alq3 (40 nm)/LiF (3 nm)/Al (130 nm) [76]. The reason for choosing rubrene as the buffer layer of the device is that its HOMO (5.4ev) value is close to that of TPD (5.5ev). As a buffer layer between the anode and the hole transport layer (HTL), the rubrene film can reduce the hole injection rate, thus making the electronic speed and hole speed of the whole device more balanced. At the same time, the introduction of buffer layer can also reduce the electrical breakdown probability of the device, which can make the device stable in a longer period of time. In conclusion, luminance intensity, efficiency and stability of OLED are improved by introducing rubrene buffer.

As a p-type material, rubrene can be used as donor layer in heterojunction OLED[77-81]. Liu et al. prepared the highly efficient blue fluorescent tandem OLED based on $C_{60}$/rubrene : $MoO_3$ as charge generation layer (CGL) [77]. By doping $MoO_3$ into rubrene, the energy levels of rubrene molecule and $C_{60}$ can be better matched, which makes the charge separation in CGL more efficient. This CGL also has good charge transport properties, which can greatly improve the carrier recombination on each emission unit. Based on the above, the maximum current efficiency and power efficiency of the device are improved, which are 43.1cd/A and 15.1 lm/W respectively. At the same time, compared with the traditional series device, its driving voltage and turn-on voltage are reduced by 6V and 2.4V respectively. Engmann et al. showed rubrene/$C_{60}$ heterojunction OLED, and probed the role of interface kinetics in the device by selective modification of the junction[78]. Wang et al. fabricated electrically driven high-performance hybrid organic/inorganic III-nitride white LEDs with Rubrene/(InGaN/GaN) multiple quantum wells (MQWs) p-n junction, and the maximum current efficiency is 15.22 cd/A [79]. This hybrid solar cell has superior long lifetime stability, which is attributed to the excellent hole transport properties of rubrene, resulting in improved charge transport and recombination. Furthermore, they also found that the main mechanism of the yellow-red luminescence is the charge capture and exciton formation in the rubrene layer. Their research provides a development direction for the study of OLED with various colors and long lifetime.

Singlet-fission（SF） induced triplet generation plays an important role in boosting the efficiency of OLED. Nagata et al. fabricated device ITO/4,4′-cyclohexylidenebis(N,N-bis(4-methylphenyl)benzenamine) (TAPC)/erbium(III) tris(8-hydroxyquinoline) (ErQ3):rubrene/2,4,6-tris(biphenyl-3-yl)-1,3,5-triazine (T2T)/2,7-bis(2,2′-bipyridine-5-yl)triphenylene (BPyTP2)/ 8-hydroxyquinolinolato-lithium (LiQ)/Mg:Ag, and its energy level diagram is shown in **Fig. 9(a)**[17]. Thermally evaporated co-deposited ErQ3 (2 mol%) : rubrene films is the emissive layer, in which rubrene was the host and SF sensitizer. The triplet excitons produced by when the singlet-fission sensitizer rubrene excited optically and electrically will be harvested by ErQ3 (**Fig. 9(b)**). In this OLED, the NIR EL peaking is at 1530 nm, and due to the SF-induced triplet generation, overall exciton production efficiency can realize 100.8%.

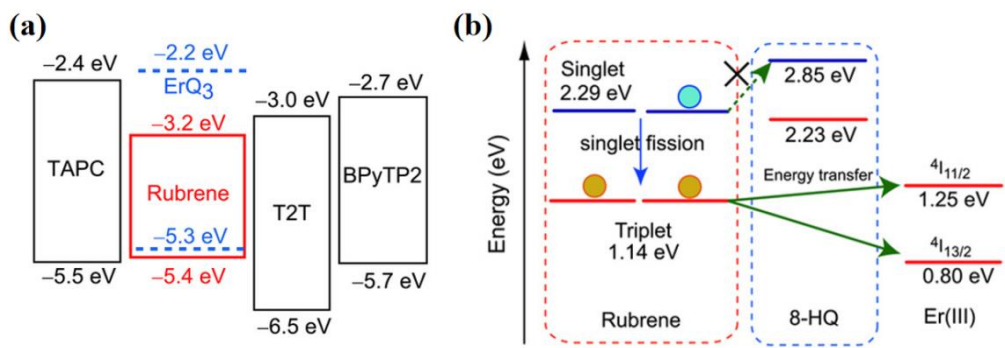

**Fig. 9** (a) Energy level diagram of the OLED. (b)The Jablonski diagram for the harvesting of triplets, which the singlet-fission sensitizer rubrene produces through singlet fission by ErQ3. **Reproduced from Ref. [17]**

**3.3 Organic solar cells/organic photovoltaic cells (OSCs / OPVs)**

As a kind of natural and environmental renewable energy, solar energy has become the focus of people's research. Both organic solar cells (OSCs) and OPV can use organic semiconductor materials to absorb photons to realize the conversion of luminous energy to electric energy. In OSCs / OPV, rubrene is often used as dopant in donor or receptor layer, which can effectively regulate the properties[82,83]. Lou et al. reported a kind of OSCs based on rubrene doped Poly(3-hexylthiphene) (P3HT):[6,6]-phenyl $C_{60}$-butyric acid methyl ester (PCBM) blends, and observed EL and photovoltaic (PV) properties simultaneously[82]. By optimizing the doping ratio of rubrene, the power convention efficiency (PV properties) and the maximum luminance (EL characteristics) can reach up to 1.05% and 65 cd/m$^2$, respectively. It provides a new route to realize the PV and EL bi-functional organic devices. Guan et al. doped rubrene into P3HT: PCBM OSCs to improve the performance[83]. Compared with undoped solar cells, when 3.5 wt% rubrene is added, the short circuit current ($J_{SC}$) and power conversion efficiency (PCE) are increased by 23.1% and 23.4% respectively, while the open-circuit voltage (Voc) and fill factor (FF) have little change. The results showed that doping rubrene will result in additional absorption and promote the charge separation, which is the important reason for the enhancement of $J_{SC}$.

Similar to OLED, rubrene can also be used in heterojunction OSCs[18,84-86]. Su et al. fabricated (ITO) / $MoO_3$ / rubrene (5,10,15,20 nm) / $C_{60}$ (30 nm) / Ag (120 nm) OPV[18]. They found that by changing the thickness of rubrene, $V_{OC}$ can be significantly increased, which is mainly caused by the singlet excitons of rubrene layer. Ndjawa et al. annealed amorphous rubrene films under different conditions, and obtained films with different structures, including amorphous films, triclinic and orthorhombic films[19]. Based on this, some rubrene/$C_{60}$ heterojunction OSCs have been fabricated to study the relationship between $E_{CT}$, which is affected by crystallinity and degree of order, and $V_{OC}$. Generally speaking, in order to make solar cells have higher energy efficiency, energy loss should be reduced as much as possible. Excessive recombination at the charge separation interface is one of the energy losses.Through measurement and calculation, it is found that the energy in CT band of rubrene crystal is comparatively low, so the loss of $V_{OC}$ is more. This is why the $V_{OC}$ yield by amorphous rubrene heterojunction OSCs were always higher than that of crystalline films. Their experiments show that the important factors of $V_{OC}$ in OSCs is the crystallinity of donor and acceptor phase，which exhibit different $E_{CT}$. Zhang et al. fabricated CuPc-$C_{60}$ planar OPV (**Fig. 10**), and introduced rubrene layer as blocking interlayer at the donor-acceptor (D-A) interface, which can effectively separate carriers[85]. Because the triplet level ($T_1$) of rubrene (1.14 ev) is similar to that of CuPc (1.16 ev), exciton transport and dissociation are permitted. The effect of interlayer on the performance of CuPc-$C_{60}$ OPV was studied by changing the thickness (0,1,2,3 nm) of rubrene. It is found that the electrons and holes are separated from the space after the interlayer is added, which reduce the saturation current density and increase carrier lifetime and $E_{CT}$. At the same time, the interlayer also lead to suppressed non-geminate recombination rate, and then increase $V_{OC}$ and power efficiency.

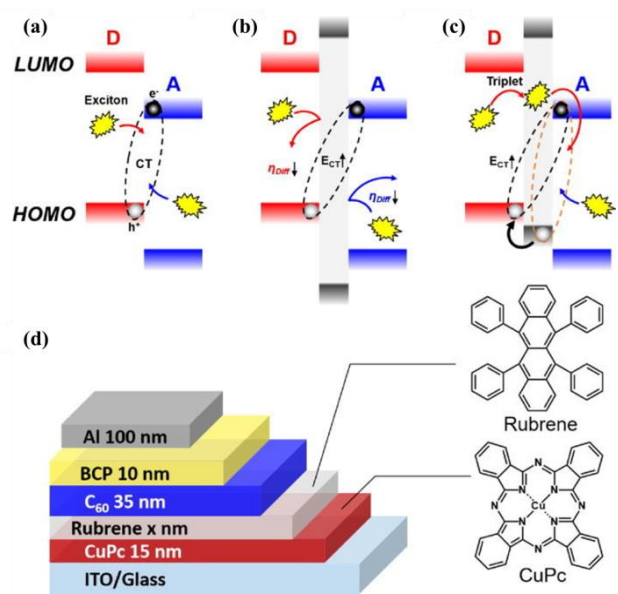

**Fig. 10** Energy diagram of the heterojunction OPV: (a) an archetypical planar heterojunction, (b) a interlayer between the donor and acceptor materials, (c) a triplet exciton permeable wide energy gap interlayer. (d) Schematic diagram of OPV device architecture. **Reproduced from Ref. [85]**

The introduction of rubrene can also improve the performance of perovskite solar cell (PSC)[87-90]. Pelicano et al. prepared a PSC containing rubrene: P3HT as hole-transport layer. The middle layer, rubrene film, covers the grain boundary of perovskite film and plays a very effective role in passivation, so the leakage current can be reduced[87]. In addition, there is a synergistic effect between rubrene layer and electron-transport layer, which greatly improves the efficiency of PSC. Cong et al. used spin coating method to prepare rubrene film, and used it as the underlayer of perovskite film to improve its crystallinity and uniform density[88]. The device is shown in **Fig. 11(a)**. As the interfacial layer between poly(3,4-ethylenedioxythiophene):poly(styrene sulfonate) (PEDOT: PSS) and perovskite, rubrene can make the hole transport easier, block the flow of electrons and align the energy leve (**Fig. 11(b)**). Finally, the PCE, $V_{OC}$, FF and stability of PSC are all improved. Qin et al. reported a novel potassium-intercalated rubrene ($K_2$Rubrene) passivation and doping material, and prepared high quality perovskite thin films by facile anti-solvent engineering[89]. The efficiency of PSC based on K$_2$rubrene is as high as 19 %.

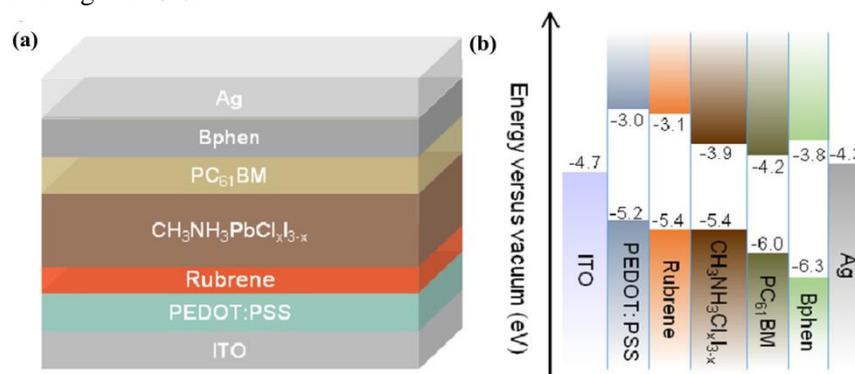

**Fig. 11** (a) Schematic diagram of perovskite solar cells device based on rubrene. (b) The energy level alignment diagram of this device .**Reproduced from Ref. [88]**

**3.4 Organic Spin-Valves(OSVs)**

Organic spintronics is an emerging discipline and organic spintronics devices combining the advantages of spintronics and organic semiconductors promise to realize transmission and storage of information with the charge and spin of electrons[91-94].With high charge mobility[48], long spin transport distance predicted hundreds of nanometers or even at the millimeter level of single-crystal rubrene in theory[95] and a spin diffusion length of 13.3 nm in amorphous rubrene thin film[21], rubrene is an ideal organic semiconductor material in the research of spintronics devices and have made significant progress[96-98].

Organic spin valves (OSVs) are the principal spintronics device and many works based on rubrene spin devices have been carried out to study injection and transport in organic materials[99,100]. Yoo et al. fabricated spin valve structure of LSMO/ LAO/ Rubrene/Fe to study the effects of bias voltage, temperature and rubrene thickness on the magnetoresistance[99] and obtained a high MR ratio of over 12% at 10 K[100]. When replacing the Fe electrode with V(TCNE) electrode, MR ratio of 4.7% at 100 K was obtained in the device [101]. Li et al. reported the spin structure of V(TCNE) /rubrene/V (TCNE), an all-organic spin valve, and they observed the MR ratio of 0.05% at 120 K using a 10 nm-thick rubrene layer[96]. Later, the author fabricated spin valve structure of Fe/rubrene/V (TCNE) spin valve and the device showing spin valve effect with a MR value of 0.01% at 300 K. They thought the small MR values may be the imperfect interface[102]. Raman et al. used SF material to inject spin-polarized carriers into rubrene and found the magnetoresistance of the junction increased as bias voltage was increased. A high value of 6% at 1.8 V for the device Al/EuS/rubrene/Fe was observed [103]. Besides using rubrene thin film as the transport medium, rubrene nanowire(~ 300 nm ) spin valve array has been reported by Alam et al, Co nanowire (~ 200 nm) and Ni thin film as ferromagnetic electrodes[22]. The spin diffusion length (L$_S$) in rubrene nanowires calculated through "modified Julliere formula" were 47.4 nm and 46.3 nm at 8 K and 100 K respectively, higher than the values in rubrene thin films.

Most organic spin valves reported have a strong temperature dependence that MR value usually decreases with increasing temperature[104,105].Therefore it is important for organic spin valves to achieve high MR response at room temperature [93,106-109]. Zhang et al. fabricated spin-valve devices with an MgO-substrate/$Fe_3O_4$/Al-O/rubrene/Co/Al stacking structure (**Fig. 12(a)**) and investigated magnetoresistance (MR) effects at room temperature and low temperature systematically based on the measurement of MR curves, current–voltage response, etc[110]. Their research results showed that the spin valve based on amorphous rubrene (2 nm) has a large MR ratio of approximately 6% at room temperature, which is one of the highest MR ratios reported so far in rubrene-based spin valves. The MR ratios were enhanced (MR ratios values up to 7.2%, 8%, and 11% for temperatures of 250, 200, and 150 K) with decreasing measurement temperatures due to the decrease of spin scattering at lower temperatures, **Fig. 12(b)**. Li et al. determined the spin diffusion length λs = 132 ± 9 nm and the spin relaxation time τs = 3.8 ± 0.5 ns in rubrene thin films at room temperature by using the inverse spin Hall effect[111]. The large $\lambda_s$ and $\tau_s$ in rubrene at room temperature explained the excellent performance of rubrene-based spin valve devices previously reported.

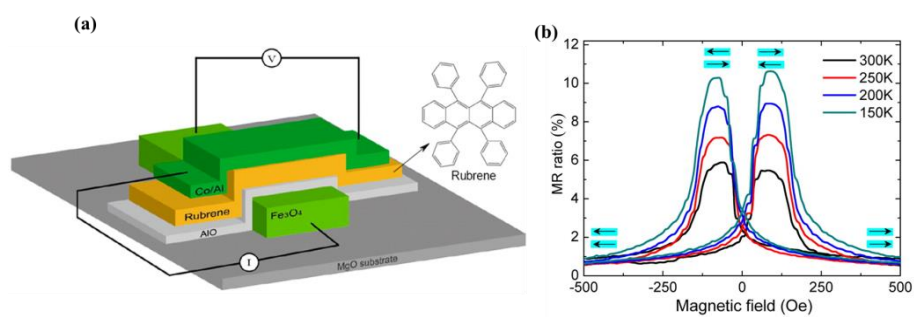

**Fig. 12** (a) Schematic view of rubrene spin-valve device. (b) Magnetoresistance curves for rubrene(2nm) organic spin valve devices measured at 300, 250, 200, and 150K, respectively. (The arrows illustrate the magnetization alignments in the antiparallel and parallel directions and the bias voltage for the measurement was fixed at 5 mV.) **Reproduced from Ref. [110]**

## 4 Summary and outlook

In summary, rubrene is a cheap, readily available, and attractive material, and its application in organic optoelectronic devices has achieved significant research progress. There are still unresolved problems in the development and application of rubrene.(1)Photo-oxidative stability. Rubrene is unstable in photo-oxygen environment[112,113] and oxidative impurities will appear during crystal growth. Interestingly, rubrene generates peroxide after photo-oxidation, but it also undergoes a reversible reaction under conditions such as heating[114]. However, it cannot be fully recovered to prepare optoelectronic devices with high purity requirements. This challenge has facilitated the development of rubrene derivatives / analogs, which was synthesized by introducing halogen or alkyl group into the side benzene ring or replacing the side benzene ring with other aromatic ring (thiophene, pyridine, etc.)[115-118]. For example, Li et al. designed and synthesized a rubrene analogue 2,2′-(6,12-bis(perfluorophenyl)tetracene-5,11-diyl)dithiophene ($SF_{10}$-RUB), exhibiting more inter-molecular interactions and lower HOMO energy levels compared with rubrene[118]. But the performance of $SF_{10}$-RUB is far inferior to rubrene, with a hole mobility of $1.73 \times 10^{-4}$ cm$^2$ V$^{-1}$ s$^{-1}$. It reflects that some rubrene analogues/derivativs are more stable than rubrene, but their photoelectric properties need to be further developed and improved. (2)Solution processing limitations. Because the crystallization process is difficult to control, the rubrene film obtained by the solution method is usually not as dense and pure as the rubrene film prepared by vacuum evaporation, and the mobility is also affected [119]. Jo et al. using spin coating prepared rubrene crystalline film doped with amorphous polymers polystyrene (PS), poly (methyl methacrylate) (PMMA), and poly (4-vinylpyridine) (P4VP), respectively[31]. Rubrene / PS films and rubrene P4VP films both exhibit an orthorhombic crystal structure , while triclinic phase presents in rubrene / PMMA blend films.The highest carrier mobility based on these mixed films is 0.52 $cm^2 V^{-1} s^{-1}$. The solution processing method is the most effective and cheapest method for realizing large-area crystal thin films, but the carrier mobility of rubrene organic thin-film transistors processed by the solution method is still relatively low. Therefore, using the solution method to obtain high-quality and large-area films and improve mobility is still the focus and difficulty of future research for rubrene. In addition, although there are many growth methods for rubrene crystalline films, the crystallization process, crystallization conditions and growth mechanism of rubrene films are not clear, and the uniformity and repeatability of the film need to be improved. Overall, rubrene is a promising materials in the field of organic optoelectronics, more attention should be paid to solve the limitations of photooxidation stability and solution processing in future research for practical applications.

## Acknowledgement

The authors acknowledge financial support from the National Key R&D Program (2016YFB0401100,2017YFA0204503)，the National Natural Science Foundation of China (51703159, 91833306, 21875158, 51633006, 51703159, 51733004).

*Nat. Mater.* 4(8), 601(2005)